\documentclass[12pt,a4paper,final]{iopart}

\usepackage{iopams} 
\usepackage{graphicx}
\usepackage{dcolumn}
\usepackage{bm}
\usepackage[breaklinks=true,colorlinks=true,linkcolor=blue,urlcolor=blue,citecolor=blue]{hyperref}

\begin{document}

\title{Ergoregion in Metamaterials Mimicking a Kerr Spacetime}

\author{D. G. Pires$^{1}$}
\address{$^1$Universidade Federal de Alagoas, Macei\'o - AL 57072-970, Brazil
}
\ead{danilo.pires@fis.ufal.br}

\author{J. C. A. Rocha$^{1}$}
\address{$^1$Universidade Federal de Alagoas, Macei\'o - AL 57072-970, Brazil
}

\author[cor1]{P. A. Brand\~ao$^{1}$}
\address{$^1$Universidade Federal de Alagoas, Macei\'o - AL 57072-970, Brazil
}

\begin{abstract}
We propose a simple singularity-free coordinate transformation that could be implemented in Maxwell's equations in order to simulate one aspect of a Kerr black hole. Kerr black holes are known to force light to rotate in a predetermined direction inside the ergoregion. By making use of cosmological analogies and the theoretical framework of transformation optics, we have designed a metamaterial that can make light behave as if it is propagating around a rotating cosmological massive body. We present numerical simulations involving incident Gaussian beams interacting with the materials to verify our predictions. The ergoregion is defined through the dispersion curve of the off-axis permittivities components.
\end{abstract}

\vspace{2pc}

\section{Introduction}

The search for new materials that can be engineered by men to perform specific useful tasks, involving light manipulation, is the major goal of photonics. From a mathematical point of view, we are actually interested in nontrivial solutions of Maxwell's equations with pre-determined material parameters that could perform specific functionalities. The study of photonic crystals, for example, allowed scientists all over the world to mimic solid state devices by employing the analogy between Schr\"{o}dinger and Maxwell's equations in periodic structures \cite{joano}. This analogy between solid state physics and optics boosted the field of photonics by incorporating new optical terms, such as photonic bandgaps, which are now engineered in the construction of photonic devices \cite{norris}. Another promising tool for dealing with the manufacturing of pre-determined functional materials has been developed in the past few years. Transformation optics (TO) provides a powerful theoretical framework for dealing with the coordinate transformation invariance of Maxwell's equations \cite{pendry,ulf,ulfbook}. Plebanski and Tamm have shown that a coordinate transformation performed on the free-space cartesian Maxwell's equations are equivalent to consider propagation of light in specific materials with properties related to the metric of the transformed space \cite{plebanski,tamm1,tamm2}. The pioneering work of Plebanski and Tamm remained practically unnoticed from the optics community until metamaterials attracted scientist's curiosity. The reason was because, in general, the implementation of a specific coordinate transformation requires anisotropic, inhomogeneous and magneto-electric media. Under some very general conditions the last requirement can be ignored. Optical cloaking, for example, from which the majority of TO applications are applied, requires only anisotropic and inhomogeneous metamaterials to turn objects invisible \cite{dehesa,ong,chen1,chen2}. A very interesting consequence of TO is that it can incorporate arbitrary spacetime transformations into its formalism to create new optical devices. As Einstein showed, general spacetime metrics actually represent gravitational fields around massive bodies. Therefore, it became possible to create metamaterials that simulate general relativity models and cosmological effects \cite{ulfart}. In particular, by choosing the correct transformation, or by direct use of the general metric, it is possible to simulate light propagating around a Schwarzschild black hole \cite{isabel,narimanov,genov,li}, to observe Hawking radiation \cite{linder}, to simulate a spinning cosmic string \cite{mackay}, optical wormhole \cite{greenleaf}, gravitational lensing \cite{tippett,cummer,crudo}, de Sitter spacetime \cite{maclat}, Big Crunch \cite{hung}, time travel \cite{boston}, all of these applications in the context of optics. 

Apart from trivial flat space, the first nontrivial exact solution of Einstein's field equations of general relativity was discovered in 1916 by Karl Schwarzschild. His solution described the gravitational field created by uncharged masses with spherical symmetry and zero angular momentum \cite{karl}. In spite of all the physical assumptions, the Schwarzschild metric is a useful approximation for slowly rotating astronomical objects. The exact solution for rotating bodies remained unsolved until 1963 when Roy Kerr analytically calculated the gravitational field around a spinning and uncharged body \cite{kerr}. With the introduction of the Kerr metric, new gravitational effects were discovered. Here we focus our attention on the ergoregion, which is a spatial region located outside a spinning black hole that forces all particles to rotate in the same direction as the hole. Unfortunately, direct implementation of the Kerr metric into the TO formalism requires magneto-electric media, which are difficult to obtain and even to perform numerical simulations \cite{comment,diean}. We propose here a simple singularity-free coordinate transformation that simulate one aspect of this exotic Kerr geometry. 

In section II we present the TO formalism necessary to a formal discussion of the metamaterials that can create a Kerr geometry along with the proposed coordinate transformation assumed throughout the paper. Section III is devoted to a short summary of the main properties of the spacetime Kerr metric to clarify what we will discuss afterwards. In section IV we present numerical simulation results of a Gaussian beam propagating around the Kerr-type black hole and discuss its main properties along with the interpretation of the ergoregion in TO.

\section{Transformation Optics}
To design a material that reproduces the ergoregion effect, light must rotate in a predetermined direction inside this region as it does around a Kerr black hole. TO allow us to design a material that, through the right choice of coordinate transformation, incoming light experiences such effect. We start by writing the vacuum Maxwell's equations in cartesian coordinates and SI units
\begin{equation}
    \label{eq1}
    \begin{array}{lll}
        \nabla \times \textbf{E} +\partial_{t}\textbf{B} &=& 0,\\
        \nabla \times \textbf{H} - \partial_{t}\textbf{D} &=& 0,\\
        \nabla \cdot \textbf{D} &=& 0,\\
        \nabla \cdot \textbf{B} &=& 0,
    \end{array}
\end{equation}
with the usual constitutive relations
\begin{equation}
    \label{eq2}
    \begin{array}{lll}
    \textbf{D}&=&\epsilon_0\overline{\varepsilon}\cdot\textbf{E},\\
    \textbf{B}&=&\mu_0\overline{\mu}\cdot\textbf{H},
    \end{array}
\end{equation}
where $\textbf{E},\textbf{H}$ are the electric and magnetic fields and $\textbf{D},\textbf{B}$ the electric displacement and magnetic induction, respectively. The permittivity and permeability tensors are denoted by an overline as $\overline{\varepsilon},\overline{\mu}$. Through a judicious choice and analogies with astrophysical bodies, we define a new transformed azimuthal coordinate $\phi'$ related to the previous original coordinate $\phi$ by
\begin{equation}
    \label{eq3}
    \phi(\rho',\phi')=\phi'+c_1 e^{-c_2\rho'}.
\end{equation}
The parameter $c_{1}$ (rad), a real number, controls the maximum value achieved by the permittivity and permeability tensors along with the twist experienced by the incident beam, meaning that large values of $c_{1}$ will increase the magnitude of $\overline{\varepsilon},\overline{\mu}$ and the rotation's strength as well [see \Eref{eq13}]. On the other hand, $c_2$ (1/m), a real nonnegative number, determines the size of the metamaterial and, consequently, the ergoregion.  Figure 1 shows the transformed space for three values of $c_{1}$, (a) positive, (b) zero and (c) negative. It should be stressed that Figure 1 is an exact numerical implementation of \Eref{eq3} and not a simple schematic representation. The choice of suppressing the axes was to improve visualization. We assume the radial coordinate $\rho$ and the $z$ coordinate to remain unchanged. We are, in fact, dealing with a two-dimensional material, and, as we shall see, this assumption restricts the analysis to the equatorial plane of the Kerr metric. Since it is desired to create an analogue of Kerr-type black holes, we apply this transformation inside the domain $\rho>R$, with $R=0.2$ m. We place an absorbing core in the region $R<0.2$ m to simulate an event horizon which is also present in Kerr black holes, where light must be absorbed. As the new coordinate $\phi'$ has dependence on $\rho$ and $\phi$, the transformation matrix can be written as
\begin{equation}
    \label{eq4}
    g_s = 
    \left[\begin{array}{lll}
    \partial \rho/\partial \rho' & \partial \rho/\partial \phi' & \partial \rho/\partial z'\\
    \partial \phi/\partial \rho' & \partial \phi/\partial \phi' & \partial \phi/\partial z'\\
    \partial z/\partial \rho' & \partial z/\partial \phi' & \partial z/\partial z'
    \end{array}\right]
    =
    \left[\begin{array}{lll}
    1 & 0 & 0 \\ 
    \Lambda_{\phi\rho'} & 1 & 0 \\ 
    0 & 0 & 1 \\  
    \end{array}\right]
    ,
\end{equation}
where  $\Lambda_{ij'} =\partial x^i/\partial x^{i'}$ with $i = \rho,\theta,\phi.$ ($x^{1} = \rho$, $x^{2} = \phi$, $x^{3} = z$). After performing the coordinate transformation, Maxwell's equations in general curvilinear coordinates take the form
\begin{equation}
    \label{eq5}
    \begin{array}{lll}
        \nabla_q \times \mathbf{\hat{E}}&=&-\partial_{t} \mathbf{\hat{B}},\\
        \nabla_q \times \mathbf{\hat{H}}&=&\partial_{t} \mathbf{\hat{D}},\\
        \nabla_q \cdot \mathbf{\hat{D}}&=&0,\\
        \nabla_q \cdot \mathbf{\hat{B}}&=&0,
    \end{array}
\end{equation}
with the new constitutive relations given by
\begin{equation}
    \label{eq6}
    \begin{array}{lll}
    \mathbf{\hat{D}}&=&\epsilon_0\hat{\overline{\varepsilon}}\cdot\mathbf{\hat{E}},\\
    \mathbf{\hat{B}}&=&\mu_0\hat{\overline{\mu}}\cdot\mathbf{\hat{H}}.    
    \end{array}
\end{equation}
Following \cite{qiu}, the material parameters and the fields ($\hat{\mathbf{E}},\hat{\mathbf{H}})$ expressed in the new coordinate system are given by
\begin{equation}
\label{eq7}
\hat{\overline{\varepsilon}}=\hat{\overline{\mu}}=\frac{\det(g_s)\det(g_{u0})}{\det(g_{u1})}P_1(g_s^T)^{-1}Q_{u_0}^{-1}g_s^{-1}Q_{u1}P_1^{-1},
\end{equation}
\begin{equation*}
\hat{\mathbf{E}}=P_1Q_{u1}^{-1}g_s(Q_{u0})^TP_0^{-1}\mathbf{E}^{i'},
\end{equation*}
\begin{equation*}
\hat{\mathbf{H}}=P_1Q_{u1}^{-1}g_s(Q_{u0})^TP_0^{-1}\mathbf{H}^{i'},
\end{equation*}
where $g_{u0}=$ diag$[1,\rho,1]$ and $g_{u1}=$ diag$[1,\rho',1]$ are the infinitesimal displacements in physical and virtual spaces, $Q_{u0}=$ diag$[1,\rho^2,1]$ and $Q_{u1}=$ diag$[1,\rho'^2,1]$ are the metrics of the physical and virtual spaces and $P_i=$ diag$[p_1^i,p_2^i,p_3^i]$ with $p_j^i=\sqrt{g_{u1_{j1}}^2+g_{u1_{j2}}^2+g_{u1_{j3}}^2}$. In cylindrical coordinates, $P_0=$ diag$[1,\rho',1]$ and $P_1=$ diag$[1,\rho,1]$. The quantities $\mathbf{E}^{i'}$ and $\mathbf{H}^{i'}$ represent the electric and magnetic field vectors expressed in the initial coordinate system. Thus, the coefficients of the material parameters expressed in the new coordinates are given by
\begin{equation}
\label{eq8}
    \begin{array}{lll}
        \overline{\varepsilon}^{\rho\rho}&=\overline{\mu}^{\rho\rho}=1,\\
        \overline{\varepsilon}^{\phi\phi}&=\overline{\mu}^{\phi\phi}=\rho^2\Lambda_{\phi\rho'}^{2}+1,\\
        \overline{\varepsilon}^{\rho\phi}&=\overline{\mu}^{\rho\phi}=\rho\Lambda_{\phi\rho'},\\
        \overline{\varepsilon}^{zz}&=\overline{\mu}^{zz}=1,
    \end{array}
\end{equation}
where $\Lambda_{\phi\rho'}=-c_1c_2e^{-c_2\rho'}$. Notice that if $\Lambda_{\phi\rho'}=0$ the transformation reduces to the identity, i.e. light travels in straight lines in the conformal cylindrical coordinates.
\begin{figure}[t]
\label{fig1}
\centering
\includegraphics[width=8cm]{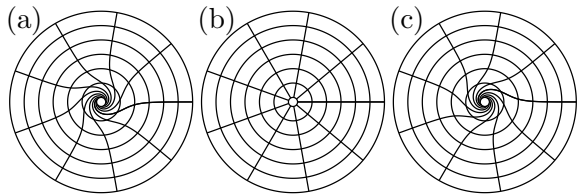}
\caption{Graphical representation of the transformation (\ref{eq3}) in cylindrical coordinate systems for (a) $c_{1} < 0$, (b) $c_{1} = 0$ and (c) $c_{1} > 0$. }
\end{figure}
The cartesian permittivities (and corresponding permeabilities $\overline{\mu} = \overline{\varepsilon}$) are easily evaluated
\begin{equation}
    \label{eq9}
    \begin{array}{lll}
        \overline{\varepsilon}^{xx}&=&\overline{\varepsilon}^{\rho \rho}\cos^2\phi +\overline{\varepsilon}^{\phi \phi} \sin^2\phi -\overline{\varepsilon}^{\phi \rho}\sin(2\theta)\\
        \overline{\varepsilon}^{xy}&=&\overline{\varepsilon}^{yx} = (\overline{\varepsilon}^{\rho \rho}-\overline{\varepsilon}^{\phi \phi})\sin\phi \cos\phi +\overline{\varepsilon}^{\phi \rho}\cos(2\theta)\\
        \overline{\varepsilon}^{yy}&=&\overline{\varepsilon}^{\rho\rho} \sin^2\phi +\overline{\varepsilon}^{\phi\phi} \cos^2\phi +\overline{\varepsilon}^{\phi \rho}\sin(2\theta)
    \end{array}
\end{equation}
with $\overline{\varepsilon}^{zz}=\overline{\mu}^{zz}=1$. We claim that the metamaterial obtained as a result of \Eref{eq9} mimicks the ergoregion, which is created by rotating black holes in the context of general relativity. The cartesian material parameters are depicted in Figure 2. The only nonvanishing components of $\overline{\varepsilon}^{ij}$ and $\overline{\mu}^{ij}$ that acquires negative values are the off-diagonal, $\overline{\varepsilon}^{xy}$  and $\overline{\mu}^{xy}$ and are zero if $c_{1} = 0$ $(\phi' = \phi)$. In section IV we will use this component to characterize the ergoregion. Parts (c) and (d) of Figure 2 depict a four-lobed pattern where each adjacent lobe alternates sign between positive and negative values. As we will see, this property is essential for the material to perform a rotation of the electromagnetic field.
\begin{figure}[t]
    \centering
    \includegraphics[width=8cm]{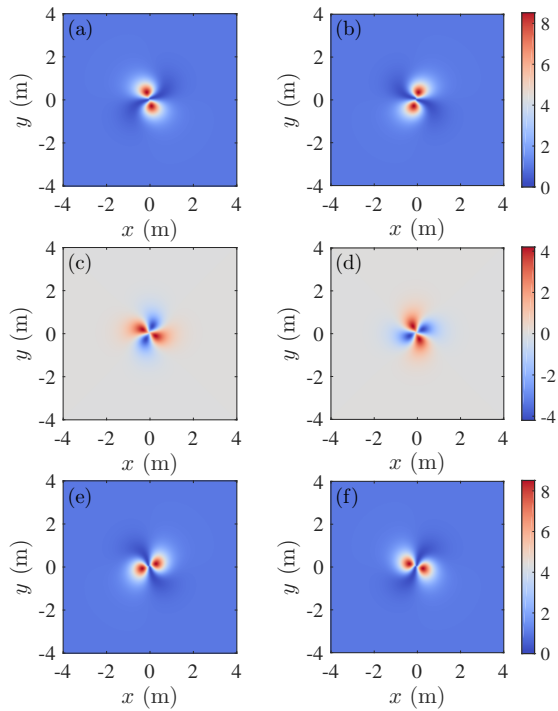}
    \caption{Permittivity and permeabillity tensors in cartesian coordinates obtained from the transformation \Eref{eq1}. (a-b) $\overline{\varepsilon}^{xx} = \overline{\mu}^{xx}$, (c-d) $\overline{\varepsilon}^{xy} = \overline{\mu}^{xy}$, and (e-f) $\varepsilon^{yy} = \overline{\mu}^{yy}$. The left column indicates materials with $c_{1} = -7$ and the right column $c_{1} = 7$. The only negative values present are due to the $xy$ crossed terms in the tensors.}
    \label{fig2}
\end{figure}

We will now assume that an electromagnetic field propagates with the electric field polarized in the $z$ direction, also called a TE mode. This means that $\mathbf{D}^{i'}$ should couple only with $\overline{\varepsilon}^{zz}$ and $\mathbf{B}^{i'}$ with $\overline{\mu}^{xx},\overline{\mu}^{xy}$ and $\overline{\mu}^{yy}$. Thus, due to the anisotropy of the medium $\mathbf{D}^{i'},\mathbf{B}^{i'}$ have different orientations compared to $\mathbf{E}^{i'},\mathbf{H}^{i'}$. The propagation equation for a TE wave, $E(x,y)$, in the equatorial plane $z=0$ can be derived from \Eref{eq5} \cite{isabel} and is given by
\begin{equation}
    \label{eq10}
    \frac{\partial}{\partial x}\left(\overline{\mu}_{xy}\frac{\partial E}{\partial y} -\overline{\mu}_{yy}\frac{\partial E}{\partial x}\right)
     -\frac{\partial}{\partial y}\left(\overline{\mu}_{xx} \frac{\partial E}{\partial y}-\overline{\mu}_{xy}\frac{\partial E}{\partial x}\right)=\omega^2\overline{\varepsilon}^{zz}E.
\end{equation}
In section IV we discuss solutions of \Eref{eq10} subject to a Gaussian beam as initial condition in order to visualize the propagation of optical beams through Kerr-type metamaterial media. However, as the optics community is not well familiarized with the Kerr black hole, we have find it useful to first review the basic concepts of the Kerr geometry in general relativity.


\begin{figure*}[t]
    \centering
    \includegraphics[width=1.00\textwidth]{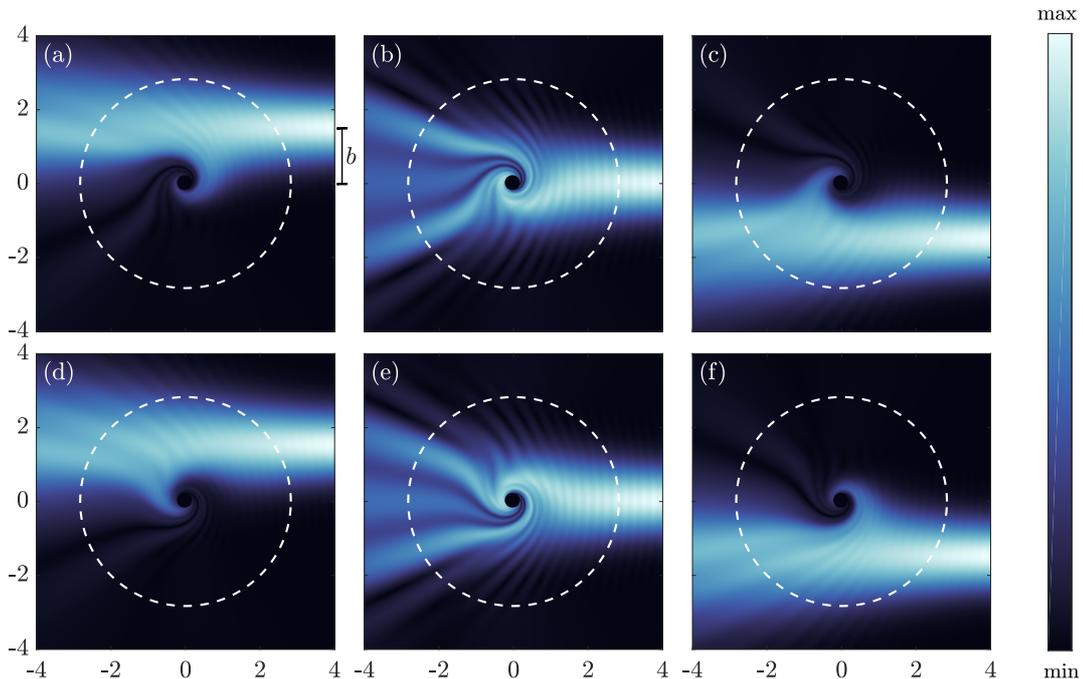}
    \caption{Numerical solution of \Eref{eq10} with incident Gaussian beams. The permittivity and permeabilitty are taken from  \Eref{eq9}. (a) Gaussian beam with initial width $0.6$ m is incident from the right to the left direction with impact parameter $b = 1.5$ and $c_1 = -7$. It is easy to see that the incident light rotates clockwise, as expected. (d) Same as part (a) but $c_{1} = 7$ indicating a counterclockwise rotation. Parts (b) and (c) are the same as part (a) with $b = 0$ and $b = -1.5$, respectively. Parts (e) and (f) are the same as part (d) with $b = 0$ and $b = -1.5$, respectively. All these six simulations are considered with $c_{2} = 3$.}

\end{figure*}



\section{Kerr Black Hole}
A brief discussion is presented to review some properties of the Kerr metric and the ergoregion radius. Rotating black holes can be described by the general spacetime metric in Boyer-Lindquist coordinates 
\begin{equation}
\label{eq11}
ds^{2} = \frac{\rho^2\Delta}{\Sigma^2}c^2dt^2-\frac{\Sigma^2\sin^2\theta}{\rho^2}(d\phi-\omega dt)^2
-\frac{\rho^2}{\Delta}dr^2-\rho^2d\theta^2,
\end{equation}
where $\Delta=r^2-2Mr+a^2$, $\rho^2=r^2+a^2\cos^2\theta$, $\Sigma=(r^2+a^2)^2-a^2\Delta\sin^2\theta$ and $\omega=g_{t\phi}/g_{\phi\phi}=2Mcra/\Sigma^2$ is the coordinate angular velocity of a zero-angular-momentum particle and the meaning of the parameters can be seen in \cite{lasenby}. The Kerr metric depends on the parameters $M$ and $a$, which refers to the black hole mass and the rotating parameter, respectively. Some properties can be deduced from the line element \Eref{eq11}. The Kerr metric is stationary and axisymmetric, i.e. it does not depend explicitly on $t$ and $\phi$. Also it is invariant under simultaneous inversion of $t$ and $\phi$, since the time reversal of a rotating object implies an object which rotates in the opposite direction. In addition, in the limits $r \to \infty$ and $a\to 0$ the metric reduces to Minkowski's metric in polar coordinates (Kerr spacetime is asymptotically flat) and to Schwarzchild metric, respectively. 

If photons are emitted in the equatorial plane ($\theta=\pi/2$) at a radial position $r$ and are going in the $\pm\phi$ direction, since $ds^2=0$ and only $dt$ and $d\phi$ do not vanish, we conclude that \cite{schutz}
\begin{equation}
\label{eq12}
\frac{d\phi}{dt}=-\frac{g_{t\phi}}{g_{\phi\phi}}\pm\left[\left(\frac{g_{t\phi}}{g_{\phi\phi}}\right)^2-\frac{g_{tt}}{g_{\phi\phi}}\right]^{1/2}.
\end{equation}
Notice that if $g_{tt}=0$, two solutions are possible: $d\phi/dt=-2\omega$ and $d\phi/dt=0$. The first solution gives $d\phi/dt$ the same sign as the parameter $a$, so the photon is sent off in the same direction as the black hole's rotation. The second solution implies that for the photon that is sent backwards, initially it does not move due the strong dragging of the orbits, so that it cannot move in the direction opposite to the rotation. So, all particles must rotate in the direction of the black hole's rotation. The spatial region where $g_{tt}=0$ is called the ergoregion and it encompasses the range from 0 to $r_{e}=M+\sqrt{M^2-a^2\cos^2\theta}$. Another way to characterize the ergoregion radius is through the effective potentials which appears in the equatorial photon motion equations in the Kerr Metric \cite{schutz,schutzart}.

It is possible to create a Kerr analogue metamaterial by direct use of the metric \Eref{eq11} as discussed in \cite{diean}.  However, this approach is much more complicated compared to the simple coordinate transformation presented in Section II. The off-diagonal term $g_{t\phi}$ creates a coupling between electric and magnetic fields \cite{isabel}, making the computational implementation considerably difficult. We propose here an easier way to simulate the ergoregion effect by using the metamaterials presented in \Eref{eq8} and we will discuss in the next section its main properties.


\section{Numerical Results and Discussion}
A close inspection of the material parameters expressed by \Eref{eq8} and \Eref{eq9} is required to validate the proposed metamaterial as simulating a Kerr geometry. Here, we present numerical simulations for the propagation of light in anisotropic inhomogeneous media which exhibits an ergoregion. The numerical validation is made by solving \Eref{eq10} in Cartesian coordinates with an incident TE polarized Gaussian beam whose beam waist is fixed at $w_0=0.6$ m and wavelength $\lambda=0.5$ m. In other words, the metamaterial may be implemented to respond very well in the regime of microwaves. We have also included the analog event horizon, that is on the same type as the Schwarzschild, by placing an absorbing circular core of radius $R = 0.2$ m at the center of the metamaterial. This core will act as the black hole's event horizon, where the strong gravity forbids light to escape. In what follows we have assumed $c_{2} = 3$ for all simulations. We choose this specific value in order to best visualize the rotational aspect of the transformation regarding the choice of $\lambda$ which fixes the computational simulation time.

Figure 3 shows the results of beam propagation through the proposed metamaterial where a Gaussian beam is incident from the right side and directed towards the localized ergoregion at the center of the figure. To observe the dragging effect we control the impact parameter $b$, which is the distance between the center of the beam and the zero on the vertical axis where the Gaussian beam is initially centered, and the sign of the transformation parameter $c_1$. For $c_1<0$ ($c_1>0$), the material mimics a black hole's rotation in the clockwise (counterclockwise) direction. Part (a) of Figure 3 shows the incident Gaussian beam with impact parameter $b = 1.5$. For this picture we chose $c_{1} = -7$ and, by inspecting Figure 1 and the coordinate transformation, we conclude that light in this particular situation must rotate clockwise. This is clearly the behavior reported in part (a). Parts (b) and (c) of Figure 3 are the same as (a) but with impact parameter  $b = 0$ and $b = -1.5$, respectively. In part (d) we have chosen $c_{1} = 7$ and, by inspecting again the coordinate transformation, we conclude that, for this particular case, light must rotate counterclockwise. This is also evident from the picture. Parts (e) and (f) of Figure 3 are the same as part (d) with $b = 0$ and $b = -1.5$, respectively. The incoming light will always rotate in the same sense as the Kerr metamaterial black hole regardless its initial direction. Figure 3 is evidence for an omnidirectional rotator for electromagnetic fields designed from an analogy with cosmological bodies.

One issue remains to be addressed. As the metamaterial described by  \Eref{eq8} is technically infinite, there is a priori no identification of a specific radius of the ergoregion. However, we have some information available directly from the materials parameters that could guide us in the characterization of the ergoregion. A glance at  \Eref{eq8} reveals that only one component is actually responsible for making light to rotate. The off-diagonal components, $\overline{\varepsilon}^{\rho\phi}$ and $\overline{\mu}^{\rho\phi}$, are the only functions not invariant under a change of sign from $c_{1}$ which, as demonstrated in Figure 3, is responsible for the sense of rotation. This creates a closer analogy with the Kerr metric, since, in the context of general relativity, off-diagonal components of the metric are responsible for the ergoregion. Therefore, the off-diagonal permittivity (and permeability) components must be mainly responsible for creating the ergoregion. In order to characterize it, we calculate the dispersion $\sigma$ of the curve
\begin{equation}
\label{eq13}
\sigma_{\overline{\varepsilon},\overline{\mu}} =\sqrt{ \frac{\int_{0}^{\infty}\rho^{2}\overline{\varepsilon}^{\rho\phi}d\rho - \left( \int_{0}^{\infty}\rho\overline{\varepsilon}^{\rho\phi}d\rho \right)^{2}}{\int_{0}^{\infty}\overline{\varepsilon}^{\rho\phi}d\rho}} = \frac{\sqrt{2}}{c_{2}},
\end{equation}
which is inversely proportional to $c_{2}$ as is intuitively expected. With the dispersion so defined we can ask what is the radius from the center of the material such that a given percentage of its bulk content is contained within it. It turns out that if we define the ergoregion radius as six times the dispersion, $r_{c} = 6\sigma_{\overline{\varepsilon},\overline{\mu}}$, then 99.8\% of the material bulk will be inside this radius. We highlight this radius in Figure 3 by a white dashed line. Figure 4 shows the plot of the coordinate transformation,  \Eref{eq3}, along with $\overline{\varepsilon}^{\rho\phi}$ and $\overline{\varepsilon}^{\phi\phi}$. The vertical line $R$ indicates the event horizon radius (absorbing material) and the vertical line $r_{e}$ the radius where 99.8\% of the metamaterial is contained. This very useful information tell us that we do not have to build an infinite material to actually perform the construction of this metamaterial in a laboratory. Also, one possibility for implementation of the ideas discussed here are liquid crystals. The implementation of liquid crystals in optical cloaking effects, for example, were already considered \cite{khoo,liu,chuang}. An alternative construction may be using hyperbolic metamaterials. Such designed materials have good performance for low frequency waves and great control of the dielectric functions for wavelengths around visible and near-infrared spectral range \cite{chen3,chen4}. We believe that the construction of the ergoregion presented in \Eref{eq9} may be possible in the bulk of liquid crystals or hyperbolic metamaterials.

It should be pointed out that metamaterials that rotates electromagnetic fields were discussed in the literature of TO in the context of optical cloaking effects and with finite annular rings metamaterials \cite{chenchan}. The approach presented here differs from previous ones by extending the transformation to `infinity' and recognizing that this effect possess a cosmological analogue.


\begin{figure}[ht]
    \label{fig4}
    \includegraphics[width=8cm]{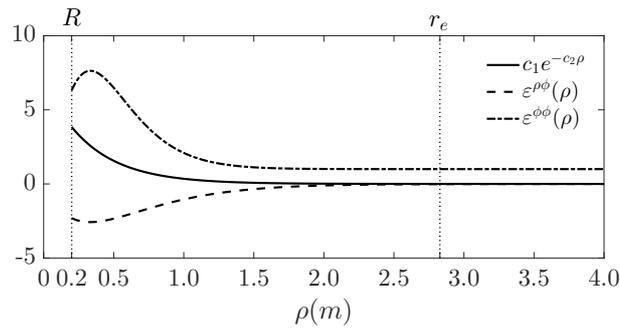}
    \centering
    \caption{In this figure, $R$ is the event horizon radius (absorbing material) and $r_{e}$ is the radius of the ergoregion defined from  \Eref{eq13}. Continuous line: coordinate transformation \Eref{eq3}. Dashed line: $\overline{\varepsilon}^{\rho\phi}$. Double dashed line: $\overline{\varepsilon}^{\phi\phi}$. They are all plotted as function of the radial coordinate with parameters $c_{1} = 7$ and $c_{2} = 3$. }
\end{figure}

\section{Conclusions}

In conclusion, based on cosmological analogies we were able to find a suitable coordinate transformation that could create a two-dimensional metamaterial mimicking the ergoregion effect, which are present in rotating Kerr black holes. We have performed numerical simulations, by solving Maxwell's equations, and characterized the response of the so called Kerr metamaterial black hole from incident Gaussian beams. The extension of the ergoregion was studied by considering its dispersion in real space.



\begin{thebibliography}{40}
\bibitem{comment} This is because the Kerr metric has non-diagonal elements in $t$ and $\phi$ Boyer-Lindquist coordinates [see  \Eref{eq11}] which requires, in the context of TO, a coupling between the electric and magnetic fields in the constitutive relations \Eref{eq6}. The metamaterial created in this way is referred to magneto-electric material.
\end{thebibliography}
\end{document}